\documentclass{svproc}
\usepackage{url}

\usepackage{graphicx}
\usepackage{subfig}
\usepackage{amsfonts, amsmath}
\usepackage{array}
\usepackage{multirow}
\usepackage{xcolor}
\usepackage{cancel}
\usepackage[linesnumbered,ruled]{algorithm2e}

\usepackage{nopageno}

\begin{document}
\mainmatter              
\title{Use of Kernel Density Estimation to understand the spatial trends of attacking possessions in rugby league}
\titlerunning{KDEs in rugby league}  
%
\author{Thomas Sawczuk\inst{1,2,3} \and Anna Palczewska\inst{1}
 \and Ben Jones \inst{2,3} \and Jan Palczewski\inst{4}
 }
\authorrunning{Thomas Sawczuk et al.} 
%
%
\institute{School of Built Environment, Engineering and Computing, Leeds Beckett University, UK,\\
\email{t.sawczuk@leedsbeckett.ac.uk}
\and
Carnegie Applied Rugby Research (CARR) Centre, Carnegie School of Sport, Leeds Beckett University, UK
\and
England Performance Unit, The Rugby Football League, Red Hall, Leeds, UK
\and
School of Mathematics, University of Leeds, UK}

\maketitle              

\begin{abstract}
Despite having the potential to provide significant insights into tactical preparations for future matches, very few studies have considered the spatial trends of team attacking possessions in rugby league. Those which have considered these trends have used grid based aggregation methods, which provide a discrete understanding of rugby league match play but may fail to provide a complete understanding of the spatial trends of attacking possessions due to the dynamic nature of the sport. In this study, we use Kernel Density Estimation (KDE) to provide a continuous understanding of the spatial trends of attacking possessions in rugby league on a team by team basis. We use the Wasserstein distance to understand the differences between teams (i.e. using all of each team's data) and within teams (i.e. using a single team's data against different opponents). Our results show that KDEs are able to provide interesting tactical insights at the between team level. Furthermore, at the within team level, the results are able to show patterns of spatial trends for attacking teams, which are present against some opponents but not others. The results could help sports practitioners to understand opposition teams' previous performances and prepare tactical strategies for matches against them.

\end{abstract}
\section{Introduction}
\label{s:intro}
Rugby league is a field based collision sport, where two teams of 13 players attempt to score points over two 40 minute halves. Points are scored by grounding the ball beyond the opposition try line or kicking the ball between the opposition posts. In the attacking phase of possession, teams attempt to progress the ball towards the opposition try line. Conversely, in the defensive phase, teams attempt to stop the opposition progressing towards their try line. Naturally, given each team has their own strengths and weaknesses, there are likely to be differences in the strategies they employ to attain maximum progress. If a practitioner is able to understand how these spatial trends of attacking possessions differ between and within teams, it could provide a significant advantage in terms of preparing tactical strategies for future matches.

One method through which the spatial trends of attacking possessions could be evaluated is Kernel Density Estimation (KDE). KDE is a non-parametric method of estimating the unknown probability density function of a dataset, which allows population-level inferences to be made based on a finite sample of data \cite{Rosenblatt1956}. By estimating the probability of a team performing an action in a given location on the pitch, it may be possible to understand the strategies employed by teams within attacking possessions. The KDE may show patterns of higher or lower density in different areas of the pitch between teams (e.g. team A may prefer to attack on the left side of the pitch, whereas team B may prefer to attack down the centre) or within teams (e.g. team A may attack more to the left against team C or perform more actions in the opposition try area against team D).

To date, no study has considered the use of KDEs to evaluate the spatial trends of attacking possessions at a team level in sport, however Mallepalle et al. \cite{Mallepalle2020} recently used the method to understand quarterback pass location distributions in the NFL. The authors used the Scott Rule of Thumb heuristic \cite{Scott2015} to identify the KDE bandwidth for each quarterback and were able to compute pass location distributions for both the league average quarterback and individual players. Using Scott's Rule of Thumb heuristic \cite{Scott2015} is likely to have resulted in different bandwidths being chosen for each player. Completing the analysis in this manner was appropriate for the work of Mallepalle et al. \cite{Mallepalle2020} as they did not intend to directly compare the KDEs obtained. However, in order to understand the differences in spatial trends of attacking possession between and within teams, our study will need to make direct comparisons between the KDEs from each subset of the data. Consequently, it would be more appropriate to remove the additional randomness caused by different bandwidths and use a single bandwidth across all KDEs in our analysis \cite{Bowman1984}.

A multi-level view of the data (like ours, which includes league, teams and opponents) was previously employed by Lichman and Smyth \cite{Lichman2014}. Their mixture-KDE method weighted individual and population levels of KDEs. The weights were identical for all individuals and were used to smooth the individual level results towards the population average to maximise predictive power. This approach was shown to be particularly useful for geolocation data with a complex geometry of the density (i.e. concentrated in corridors around roads) and a low number of observations per individual. However, it is unlikely to benefit our analysis as the location of actions in rugby is much more evenly spread across the field and we have a sufficient number of observations at each level of analysis.

Langlois et al. \cite{Langlois2012} provide one of few examples of studies which have explicitly compared KDEs between subsets of a dataset, in their case the length-frequencies of different species of fish. The authors used the Sheather-Jones method \cite{Sheather1991} to estimate the bandwidth for each species of fish's KDE before taking the geometric mean of these bandwidths. The KDEs were re-run at this geometric mean bandwidth and compared using a statistical test comparing expected vs actual areas of difference between the length-frequencies of the fish calculated using different methods. The work of Langlois et al. \cite{Langlois2012} provides a starting point through which the spatial trends of rugby league teams' attacking possessions could be considered by using the geometric mean of their KDE bandwidths. However, rather than solely considering whether there is a difference between teams' possession densities, it is important to gain understanding of how large the difference is in terms of size and location on the pitch. These insights are provided by the Wasserstein distance, which accounts for both the differences in the probability density and how those differences are spread on the pitch. The latter feature is particularly useful in analysing possession densities where large spatial differences could be linked to differences in team strategies.

In rugby league, the closest study to our work is that of Sawczuk et al. \cite{Sawczuk2021}, who attempted to understand the spatial trends of attacking performances by computing the expected points gained by an action conditional on its location. They used a Markov Decision Process with fixed grid sizes, and z-score analysis to identify differences between teams. Our work differs from theirs as it solely considers how likely an action is to occur in a given location, not what reward is likely to be obtained by it. Furthermore, we provide a continuous representation of the pitch, rather than using a discrete grid, which provides the model with much more flexibility to understand the differences present in a dynamic team sport like rugby league.

In this study we evaluate the spatial trends of attacking possessions in rugby league for the first time by estimating the 2-dimensional probability density function of a team's actions across a rugby league pitch. We use KDEs to estimate the probability density function for the whole league, for each team overall and then for each opponent each individual team faced over the duration of the 2021 Super League season. We use the Wasserstein distance to establish the size of differences between and within teams' attacking KDEs. Finally, we use KDE plots to identify those areas where teams' attacking densities differ and what tactical insights these differences may bring.

The paper is organised as follows: Section \ref{s:methodology} introduces kernel density estimation, bandwidth selection, and the Wasserstein distance between distributions; Section \ref{s:data} describes the dataset used and preprocessing steps taken; Section \ref{s:kde_rugbyleague} describes the KDE approach to evaluating the spatial trends of attacking possessions in rugby league; Section \ref{s:results} discusses the results; and Section \ref{s:conclusion} concludes and outlines future research directions.

\section{Methodology}
\label{s:methodology}
In this section we outline the KDE procedure and Wasserstein distance metric used within this study. We recall the definition of KDE as a non-parametric method of estimating the unknown probability density function of a dataset \cite{Rosenblatt1956}. The bivariate KDE ($\hat{f}_H$) of a sample of $2$-dimensional vectors $(\vec{x}_i)_{i=1}^n$, in this case a vector containing the $x$ and $y$ co-ordinates of action locations, is defined as
\begin{equation}
    \hat{f}_H(\vec{x})= \frac{1}{n} \sum_{i=1}^n{K_H(\vec{x} - \vec{x}_i)},
    \label{eq:kde}
\end{equation}
where $K$ is the kernel and $H$ is a $2 \times 2$ smoothing matrix. For simplicity, we use a diagonal smoothing matrix with the same smoothing constant $h$ (the bandwidth) for both directions. We employ the standard bivariate normal kernel
\begin{equation}
    K_h(\vec{x}) = (2\pi h)^{-1} e^{-\frac{\|\vec{x}\|^2}{2h}},
    \label{eq:mvnorm}
\end{equation}
where $\|\vec{x}\|$ is the Euclidean norm of $\vec{x}$.

The most important free parameter within KDE is the bandwidth. This parameter influences the smoothness of the KDE model and controls ``overfitting''. Smaller bandwidth values result in a more jagged appearance with larger peaks and troughs, whereas larger values result in a much smoother appearance with smaller peaks. Typically, as the size of the dataset increases, the optimal bandwidth size reduces \cite{Zambom2012}. However, when comparing KDEs it is important to use the same bandwidth to avoid observing the artificial differences that can be induced by different bandwidth sizes \cite{Bowman1984}.

Given its importance to the estimates provided by KDE, numerous methods have been identified through which the optimal bandwidth can be selected, including visual inspection, fitting to a reference distribution \cite{Silverman1986,Scott2015}, estimation to minimise the mean integrated square error \cite{Park1990,Sheather1991} and cross-validation \cite{Stone1984,Bowman1984,Marron1989}. Contrary to the assumptions underlying these approaches, our rugby league data is auto-correlated as locations of consecutive actions in a play are not independent. The effect of auto-correlation on optimal bandwidth selection is shown in \cite{Fleming2015} in the framework of minimising the mean integrated square error for a Gaussian reference distribution. Their method performs particularly well in the study of highly correlated animal location data. However, the stretches of correlated locations in our dataset are relatively short and of varying length, so the approach in \cite{Fleming2015} with a reference Gaussian distribution would not offer an improvement over a cross-validation approach which is distribution-free. Furthermore, it has been suggested that the cross-validation approach is more responsive to the different sample sizes we will encounter within this study \cite{Zambom2012}. We therefore adopt cross-validation as our method of optimal bandwidth selection.

The $p$-Wasserstein distance between two distributions $\mu$ and $\nu$ is calculated as \cite{Dobrushin,Panaretos2019}:
\begin{equation}
W_p(\mu,\nu) = \Big(\inf_{\gamma \in \Gamma(\mu, \nu)} \int_{M \times M} \|x - y\|_p^p \gamma(dx, dy) \Big)^{\frac{1}{p}},
    \label{eq:wasserstein}
\end{equation}
where $M$ is a 2-dimensional space of coordinates on the pitch, $p \ge 1$, $\| (x, y) \|_p = (|x|^p + |y|^p)^{1/p}$ is the $L_p$ norm\footnote{$L_1$ norm is sometimes called Manhattan norm and $L_2$ norm is the Euclidean norm.} and $\Gamma(\mu, \nu)$ is the set of all couplings of distributions $\mu$ and $\nu$, i.e., the set of all joint distributions on $M \times M$ with marginals $\mu$ and $\nu$. 

\section{Data}
\label{s:data}
\begin{figure}[htb!]
\centering
\includegraphics[scale=0.4]{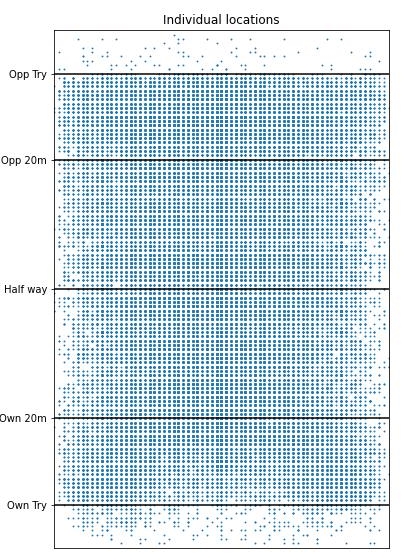}
\caption{Location of 99,966 actions used in this study. Dots representative of location only; quantity of actions at each location is not displayed in this plot.}
\label{fig:data_locations}
\end{figure}

The dataset used for this study was provided by StatsPerform and includes 138 matches (contested by 12 teams) from the 2021 Super League season, downloaded from www.optaprorugby.com. StatsPerform provides details of a multitude of different events, including passes made, ball receptions, tackles and video referee events. In order to evaluate the spatial trends of attacking possessions of rugby league teams, only events related to attacking actions were required. This is because only those actions where a player was in possession of the ball could result in a point scoring action occurring. Consequently, after removing the actions not relevant to this study (including defensive, auxiliary and off-the-ball events, actions where players failed to maintain control of the ball (e.g. dropped catches) and any duplicate location actions (e.g. when the player is coded as catching the ball and running with it from the same location), we reduced the dataset of 557,050 actions into a dataset of 99,966 actions. For each action, only the continuous $x,y$ location denoting its position relative to a standardised 100m x 70m rugby league pitch was considered. In the $x$-direction, values range from $0$ to $70$; in the $y$-direction values range from $-10$ to $110$. Actions occurring outside the 0-100m pitch are located in the attacking team's try area (0 to $-10$m) or the opposition team's try area (100 to 110m). Table \ref{t:sample_possession} provides a sample possession from the dataset. Figure \ref{fig:data_locations} shows locations of the 99,966 actions used in this study.

To enable the comparison of spatial trends of attacking possessions between and within teams, the 99,966 action dataset was split into a total of 144 subsets. To study individual teams' overall attacking trends, the data were split into 12 subsets. Each of these subsets contained actions for when the attacking team was in possession of the ball, irrespective of the opposition (i.e. it included data from matches of the attacking team against all opponents). To study the differences within individual teams, we split the dataset into 132 subsets to evaluate every opposition team individually. Each of these subsets only contained actions for when the attacking team was in control of the ball against a specific opponent. Table \ref{t:descriptives} provides descriptive data for the action counts of the subsets.

\begin{table}[bt]
\caption{Sample possession within the dataset. Actions were not analysed for the KDEs but are included for descriptive purposes.}
\label{t:sample_possession}
\begin{center}
\begin{tabular}{ccccc}
\hline
Attacking Team & Defending Team & $x$ & $y$ & Action\\
\hline
St Helens & Salford & 9 & 4 & Catch\\
St Helens & Salford & 9 & 6 & Run\\
St Helens & Salford & 14 & 11 & Pass\\
St Helens & Salford & 22 & 13 & Pass\\
St Helens & Salford & 12 & 12 & Run\\
St Helens & Salford & 37 & 16 & Run\\
St Helens & Salford & 36 & 24 & Run\\
St Helens & Salford & 54 & 35 & Kick\\
\hline
\end{tabular}
\end{center}
\end{table}

\begin{table}[bt]
\caption{Descriptive data for action counts in the data subsets}
\label{t:descriptives}
\begin{center}
\begin{tabular}{ccccc}
\hline
Comparison & Subsets & Median & Interquartile Range & Minimum/Maximum\\
\hline
Between team & 12 & 8105 & 7596-8937 & 7203/10324\\
Within team & 132 & 732 & 622-858 & 277/1551\\
\hline
\end{tabular}
\end{center}
\end{table}

\section{Use of KDEs to identify the spatial trends of attacking possessions in rugby league}
\label{s:kde_rugbyleague}

In this study, we attempt to identify and evaluate the spatial trends of attacking possessions in rugby league. To understand these trends, we first create 145 2-dimensional KDEs using data from the whole league (1 KDE using all 99,966 actions; whole league KDE), each teams' overall attacking data (12 KDEs; team overall KDE) and each teams' data against each individual opponent (132 KDEs; team-opponent KDE). The team KDEs were calculated from the subsets of data identified in Section \ref{s:data}.

The bandwidth for the KDEs is chosen via a two stage process. First, the optimal bandwidths for the 132 subsets on the team-vs-opponent level are identified using 10-fold cross validation via the scikit learn package \cite{scikit-learn} in Python. Second, the geometric mean of those bandwidths is calculated; it resulted in the bandwidth equal to 4.10. We then re-run all 145 KDEs using the selected bandwidth 4.10 \cite{Bowman1984}.

We use the 1-Wasserstein distance (i.e., $p=1$ in \eqref{eq:wasserstein}) to globally evaluate the differences in trends of attacking possessions between and within teams. First, to understand the between-team differences in the spatial trends of attacking possessions, the Wasserstein distance between the whole league KDE and each team's overall KDE is considered. These distances identify how similar the location of a team's attacking actions are compared to the league average. A smaller distance indicates greater similarity to the league average. Next, we consider the within team differences by comparing the team's overall KDE with their team-opponent KDEs. We evaluate these results in two ways. Firstly, we group the Wasserstein distances by the attacking team and compare the Wasserstein distances against each opponent to establish how much the attacking team's spatial trends vary between opponents. Secondly, we group together the results by a specified opponent and compare all attacking teams' Wasserstein against that team. The Wasserstein distances for the attacking teams are still calculated using the attacking team's overall KDE. Grouping by a specified opponent in this manner allows us to identify which opponents are associated with greater or less variability in the spatial trends of the teams attacking against them. All Wasserstein distances were calculated using the POT package \cite{pythonot} in Python, using the algorithm described by Bonneel et al. \cite{Bonneel2011}.

\section{Results}
\label{s:results}

\begin{figure}[htb!]
\centering
\includegraphics[scale=0.5]{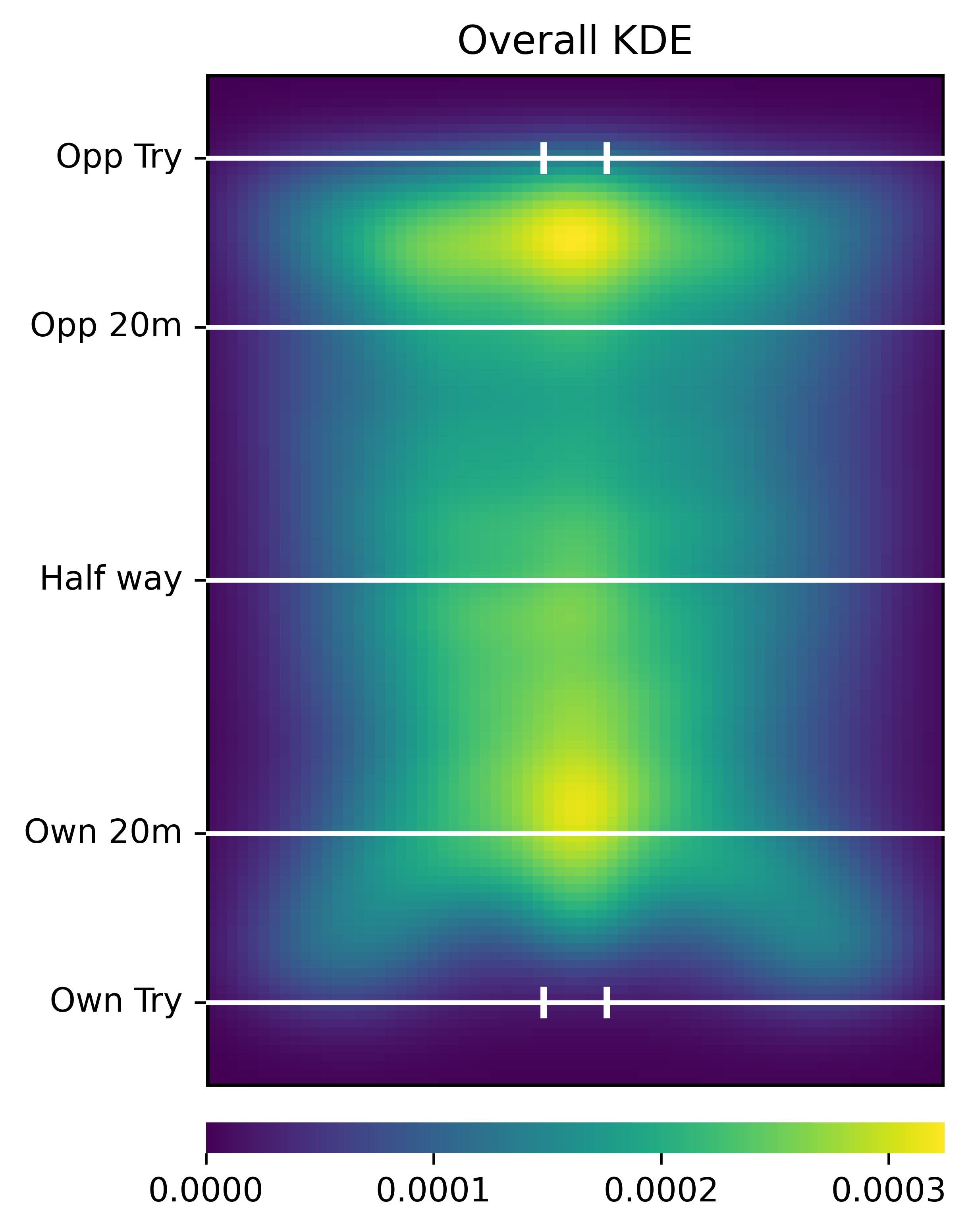}
\caption{Kernel Density Estimation plot for whole league data at bandwidth 4.10}
\label{fig:overall_kde}
\end{figure}

Figure \ref{fig:overall_kde} displays the whole league attacking KDE. There is a small preference across the whole league to perform actions in centre-left locations across the length of the pitch. Likewise there are greater densities of actions on the team's own 20m line and approximately 10m from the opposition try line. Table \ref{t:bet_wit_teams} provides 1-Wasserstein distances at the between and within-team level for all teams. At the between-team level, these values represent the difference between the whole league KDE (Figure \ref{fig:overall_kde}) and the team's overall KDE. At the within-team level, they represent the difference between the team's overall attacking KDE and the team-opponent KDE.

\begin{table}[bt!]
\caption{Between and within team 1-Wasserstein distances for spatial trends of attacking possessions. Row represents the attacking team, column represents the individual opponent. 'All' column provides 1-Wasserstein distances between the team's overall KDE and the whole league KDE. Columns 'Cas' to 'Wig' provide 1-Wasserstein distances between the team-opponent KDE and the team's overall KDE. At the row level, means and standard deviations do not include the 'All' column values. Team names are abbreviated for space.}
\label{t:bet_wit_teams}
\begin{center}
\begin{tabular}{c|c|cccccccccccc|cc}
\hline
Team & All & Cas & Cat & Hud & Hul & HKR & Lee & Lei & Sal & StH & Wak & War & Wig & Mean & SD\\
\hline
Cas & 3.07 & & 4.02 & 6.64 & 4.74 & 5.19 & 5.36 & 4.29 & 3.38 & 5.48 & 2.48 & 2.88 & 3.17 & 4.33 & 1.28\\
Cat & 1.08 & 6.21 & & 3.00 & 2.76 & 3.17 & 10.16 & 4.59 & 8.27 & 6.63 & 1.87 & 4.27 & 4.85 & 5.07 & 2.54\\
Hud & 3.38 & 4.45 & 1.69 & & 3.85 & 2.77 & 2.77 & 3.99 & 2.24 & 11.90 & 4.43 & 2.96 & 5.21 & 4.20 & 2.76\\
Hul & 1.38 & 2.75 & 9.30 & 4.79 & & 4.34 & 2.55 & 4.19 & 4.63 & 6.83 & 4.12 & 7.61 & 4.05 & 5.01 & 2.06\\
HKR & 2.12 & 7.04 & 5.86 & 3.32 & 6.77 & & 2.00 & 4.76 & 4.27 & 10.58 & 2.82 & 6.19 & 2.68 & 5.12 & 2.51\\
Lee & 2.27 & 5.32 & 7.22 & 4.68 & 3.35 & 3.06 & & 7.49 & 3.82 & 6.28 & 2.63 & 5.24 & 7.74 & 5.17 & 1.84\\
Lei & 2.61 & 15.00 & 6.31 & 3.10 & 9.56 & 2.26 & 6.05 & & 3.39 & 3.86 & 3.72 & 4.86 & 3.47 & 5.60 & 3.73\\
Sal & 2.36 & 5.51 & 2.83 & 7.11 & 4.31 & 3.63 & 5.03 & 2.09 & & 5.20 & 7.03 & 5.49 & 3.87 & 4.74 & 1.59\\
StH & 1.60 & 5.47 & 6.11 & 4.63 & 3.71 & 7.37 & 4.14 & 2.73 & 2.83 & & 1.86 & 2.61 & 3.91 & 4.12 & 1.67\\
Wak & 1.85 & 4.43 & 4.93 & 3.16 & 4.81 & 3.08 & 4.84 & 6.25 & 5.25 & 3.28 & & 7.08 & 3.09 & 4.56 & 1.34\\
War & 1.72 & 3.72 & 9.21 & 2.56 & 4.46 & 5.08 & 2.50 & 3.48 & 3.03 & 4.57 & 5.45 & & 4.28 & 4.40 & 1.87\\
Wig & 1.34 & 3.60 & 10.56 & 2.02 & 3.27 & 3.38 & 4.74 & 4.49 & 5.70 & 6.26 & 2.20 & 3.91 & & 4.56 & 2.38\\
\hline
Mean & 2.07 & 5.77 & 6.19 & 4.09 & 4.69 & 3.94 & 4.56 & 4.40 & 4.26 & 6.44 & 3.51 & 4.83 & 4.21\\
SD & 0.71 & 3.30 & 2.77 & 1.64 & 1.94 & 1.47 & 2.30 & 1.49 & 1.69 & 2.64 & 1.64 & 1.69 & 1.39\\
\hline
\end{tabular}
\end{center}
\end{table}

\begin{figure}[hbt!]
\centering
\includegraphics[scale=0.45, angle=90]{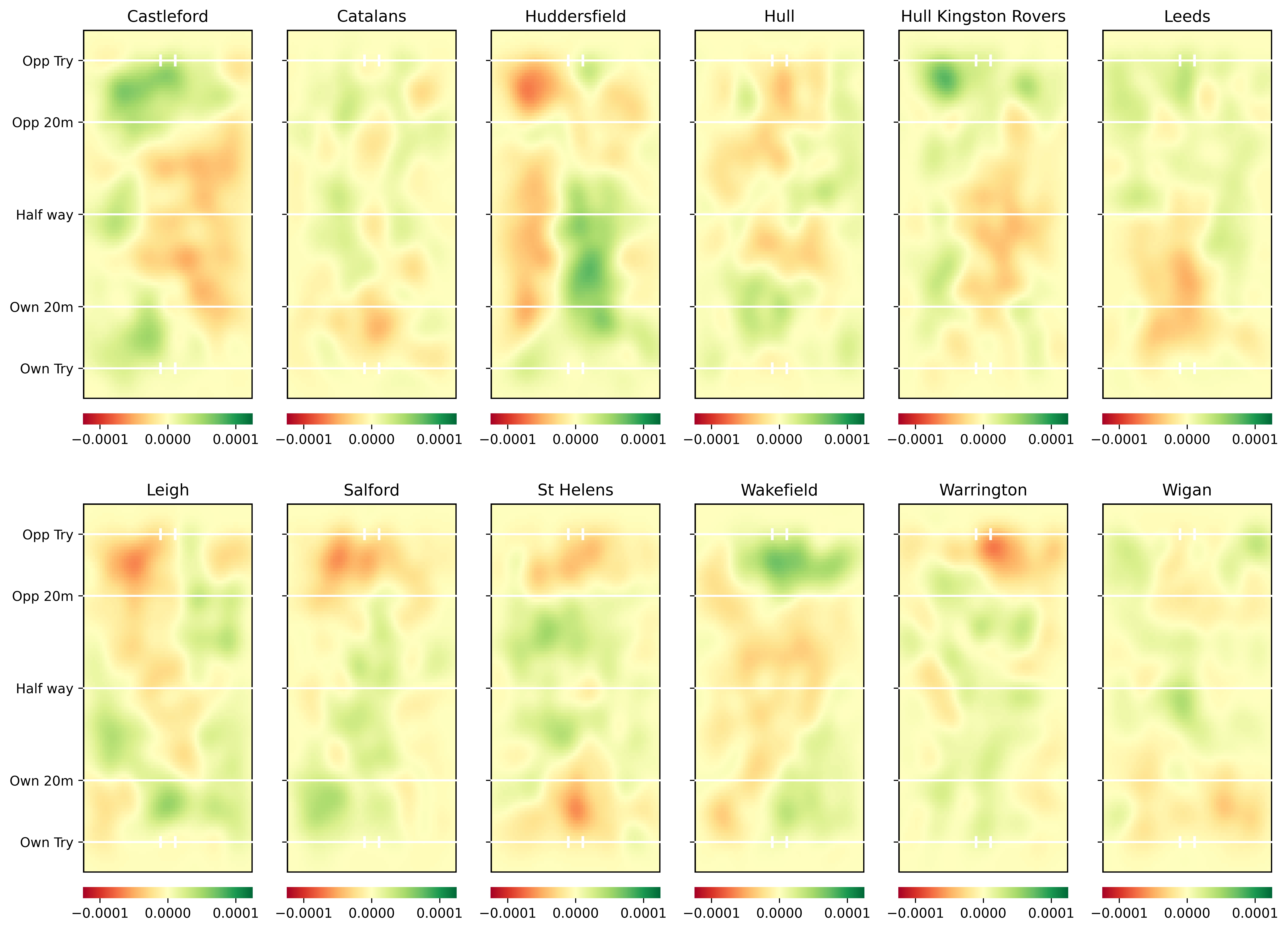}
\caption{Comparison between whole league attacking KDE and each teams' overall KDE. Green areas are areas represent areas where the team has a higher density than the whole league, red areas have lower densities.}
\label{fig:attacking_diffs}
\end{figure}

Between-team differences in the spatial trends of attacking possessions are shown via the 'All' column in Table \ref{t:bet_wit_teams}. These values represent the differences between each team's overall KDE and the whole league KDE. They provide two noteworthy findings. Firstly, the two teams with the highest Wasserstein distances are Castleford (Cas) and Huddersfield (Hud). Visual inspection of the KDEs for these two teams (Figure \ref{fig:attacking_diffs}) shows some clear tactical insight for how both teams attack. Castleford have much higher densities on the left side of the pitch than the whole league model across virtually the length of the pitch. Huddersfield on the other hand have lower densities on the left wing, but have a very high build up of density in centre-right locations between their own 20m and the half-way line. Secondly, of the four teams with the lowest Wasserstein differences, three finished in the top 4 positions in the league (Catalans (Cat) 1st, St Helens (StH) 2nd and Wigan (Wig) 4th). Visual inspection of these teams' plots in Figure \ref{fig:attacking_diffs} shows much paler colours indicating that these teams had weaker preferences with regards to their attacking locations and had a more uniform distribution of actions across the pitch than Castleford or Huddersfield. Given the finishing positions of Castleford (8th) and Huddersfield (7th), it is possible that opposition teams were able to prepare and defend against their spatial preferences much better than those teams who were more likely to perform actions across the width of the pitch.

\begin{figure}[hbt]
\centering
\includegraphics[scale=0.35]{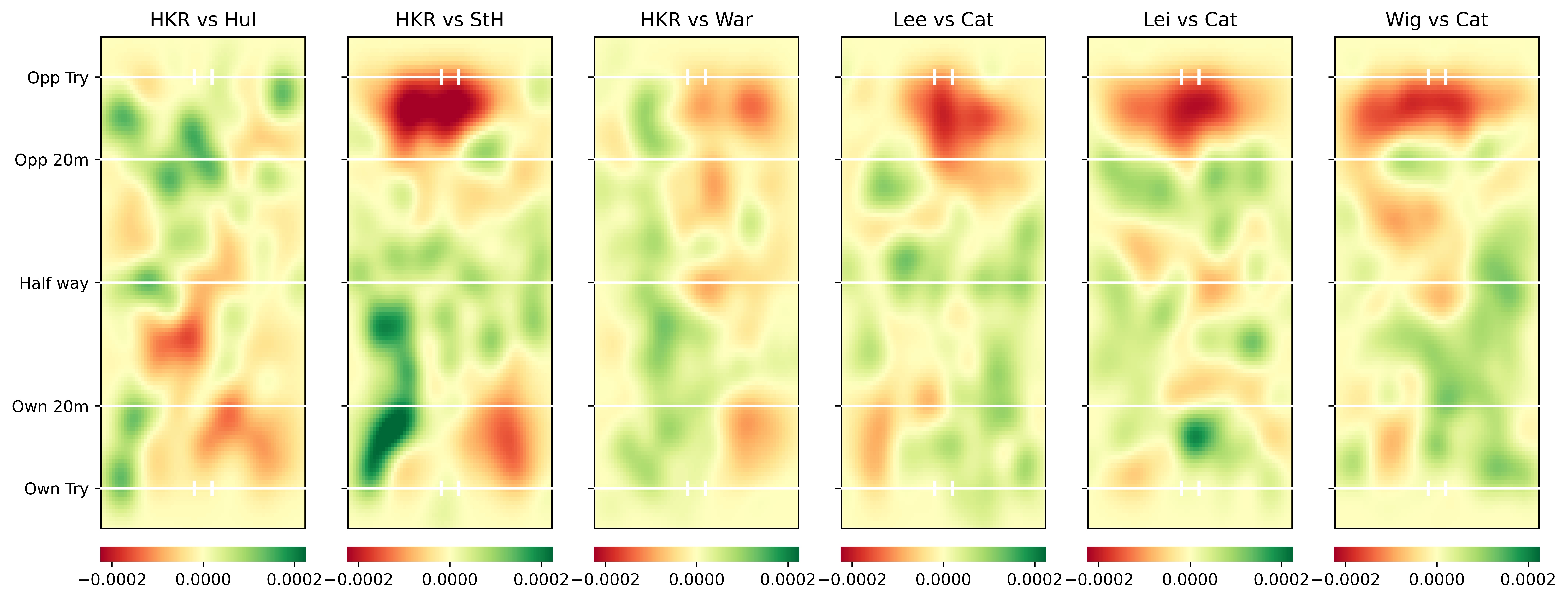}
\caption{Six plots highlighting the differences that can be seen at the within-team level. Left three: Hull Kingston Rovers (HKR) showing higher densities on the left side of the pitch against Hull (Hul), St Helens (StH) and Warrington (War). Plots show differences between HKR overall KDE and HKR-opponent KDE. Right three: KDE plots for Leeds (Lee), Leigh (Lei) and Wigan (Wig) attacking against Catalans (Cat). Plots show differences between the attacking team's overall KDE and the attacking team vs Catalans KDE.}
\label{fig:specific_diffs}
\end{figure}

The Wasserstein distances against individual opponents in Table \ref{t:bet_wit_teams} provide information regarding the within-team variability in spatial attacking trends. At the row level, the variability in KDEs for each attacking team is aggregated using the mean and standard deviation. Alone, this provides limited information, although the team with the highest mean and standard deviation (Lei: Leigh) finished bottom of the league potentially indicating they were least able to attack in a consistent manner between opponents. However, when considering the four highest Wasserstein distances for each team, and using within-team KDE difference plots, greater insights can be obtained. For example, it can be seen that Hull Kingston Rovers (HKR) may have a slight left wing preference as this was prominent in their matches against Hull (Hul) and Warrington (War). Similarly, a large amount of the spatial differences for Catalans (Cat) appear to be related to their densities within the opposition 20m. At the column level, the Wasserstein distances in Table \ref{t:bet_wit_teams} help to understand the within-team differences that are present when all teams attack against a specific opponent. These differences are particularly insightful at a mean level as they show that Catalans (Cat) and St Helens (StH) have higher mean Wasserstein distances than all other defending teams. Visual inspection of the within-team plots of all attacking teams against these two opponents shows that the majority of this difference generally appears to be accounted for by the attacking team's lower densities within the opposition 20m. Given Catalans and St Helens were the two best performing teams in the league, their ability to reduce the opposition's probability of performing an action within their 20m may not be surprising, but is certainly insightful when visualised. Figure \ref{fig:specific_diffs} shows some of the differences that can be depicted at the within-team level.

\section{Conclusion}
\label{s:conclusion}
This study is the first to use KDEs to model the spatial trends of attacking possessions in rugby league. We show how this model is able to differentiate between teams' attacking trends when considering all the team's actions across the season. We also highlight specific instances where it is able to highlight these differences at a within-team level. Future studies may wish to consider different methods of assessing the differences between KDEs, particularly those methods which are able to link these differences to tactical insights similar to this study. 

\section{Acknowledgment}
This was a cross-school project involving the School of Built Environment, Engineering and Computing, Carnegie School of Sport and Rugby Football League, who provided access to the data. The authors express their gratitude to all parties involved.

\bibliographystyle{splncs03}     
\bibliography{references}  

\end{document}